\documentclass[aip]{revtex4-1}
\usepackage{graphicx}
\usepackage{amsfonts}
\begin{document}
\title{Stretched exponential behavior and random walks on diluted
  hypercubic lattices}

\author{N. Lemke} \email{lemke@ibb.unesp.br}
\affiliation{ 
Departamento de F\'{\i}sica e Biof\'{\i}sica\\
Instituto de Bioci\^encias de Botucatu\\
UNESP - Univ Estadual Paulista\\
Distrito de Rubi\~ao Jr. s/n \\
Botucatu, S\~ao Paulo, 18618-970, Brazil
}

\author{Ian A. Campbell} \email{Ian.Campbell@univ-montp2.fr}
\affiliation{
  Laboratoire Charles Coulomb,\\
  Universit\'e Montpellier II, 34095 Montpellier, France }

\date{Received: date / Revised version: date}

\begin{abstract}
  Diffusion on a diluted hypercube has been proposed as a model for glassy
  relaxation and is an example of the more general class of stochastic
  processes on graphs.  In this article we determine numerically
  through large scale simulations the eigenvalue spectra for this
  stochastic process and calculate explicitly the time evolution for
  the autocorrelation function and for the return probability, all at
  criticality, with hypercube dimensions $N$ up to $N=28$. We show
  that at long times both relaxation functions can be described by
  stretched exponentials with exponent $1/3$ and a characteristic
  relaxation time which grows exponentially with dimension $N$. The
  numerical eigenvalue spectra are consistent with analytic
  predictions for a generic sparse network model.

\end{abstract}
\maketitle
\section*{Introduction}

In 1854 R. Kohlrausch used a phenomenological expression
\begin{equation}
  \label{kohl}
  q_{K}(t)=\exp(-(t/\tau)^\beta)
\end{equation}
to parametrize the non-exponential decay of the electric polarization
of Leyden jars (primitive capacitors)\cite{RK}; his son F. Kohlrausch
later used the same expression to analyse creep in galvanometer
suspensions \cite{FK}. A century later, in 1951 Weibull introduced
\cite{weibull} the closely related Weibull function; this survival
probability function \cite{eliazar} which is widely used in the
engineering literature is strictly of the Kohlrausch form,
Eqn. (\ref{kohl}). In 1970 Williams and Watts re-discovered the
Kohlrausch function in the context of dielectric
relaxation\cite{WW}. Under the name of ``stretched exponential''
\cite{chamberlin} the KWW (Kohlrausch-Williams-Watts) function has
become ubiquitous in phenomenological analyses of non-exponential
relaxation data, experimental or numerical.  In particular the KWW
form was used by Ogielski in a phenomenological fit the decay of the
autocorrelation function at equilibrium for a $3d$ Ising spin glass
model \cite{ogi1985}.

Many arguments have been given as to why under certain assumptions,
specific systems should show KWW relaxation
\cite{phil,havl,rasa,dons1,dons2,gras,gotz,ian1,ian2,ian3}, but there
have always been lingering suspicions that for most cases the KWW
expression is nothing more than a convenient fitting function of no
fundamental significance.

It was conjectured \cite{ian1} that KWW relaxation is the signature of
a complex configuration space. Thus from the argument which follows it
was suggested that random walks on a diluted hypercube (a hypercube
with a fraction $p$ of vertices occupied at random) near the critical
concentration for percolation $p_{c}$ \cite{erdos1979} would lead to
an autocorrelation function decay of the form $q(t) \sim
\exp[-(t/\tau)^{\beta}]$, with a specific value of the exponent,
$\beta = 1/3$.

For random walks at percolation threshold in a randomly occupied
Euclidean (flat) space of dimension $d$ such as $\mathbb{Z}^{d}$, the
familiar Fickian diffusion law $\langle R^2\rangle \sim t$ is replaced
by a sub-linear diffusion $\langle R^2\rangle \sim t^{\beta_{d}}$,
with $\beta_{d} \equiv 1/3$ for $d\geq 6$ \cite{alexander:82}. Random
walks on the surface of a full [hyper]sphere $\mathbb{S}_{d-1}$ in any
dimension $d$ are characterized by the generic law $\langle
\cos(\theta)\rangle = \exp(-(t/\tau))$ where $\theta$ denotes the
generalized angular displacement of the walker
\cite{debye,caillol}. It was argued \cite{ian1} that random walks on
percolation clusters at threshold inscribed on [hyper]spheres would be
characterized by relaxation of the form $\langle \cos(\theta)\rangle =
\exp(-(t/\tau)^{\beta_{d}})$ with the same exponents $\beta_{d}$ as in
the corresponding Euclidean space. This was demonstrated numerically
for $d = 3$ to $8$ \cite{jund}. A hypercube being topologically
equivalent to a hypersphere, for random walks on a diluted hypercube
at threshold one then expects stretched exponential relaxation with
exponent $\beta = 1/3$.

The diluted hypercube at threshold can alternatively be considered as
a specific example of a sparse graph. Remarkably, analytic expressions
for diffusion on general sparse graphs \cite{bray1988,samukhin2008} derived
from a quite different line of argument also lead to stretched
exponential relaxation expressions with the same specific value
$\beta=1/3$ for the exponent.

Here we present numerical data for random walks on the diluted
hypercube at threshold up to dimension $N=28$ which are consistent
with these conclusions.
% prediction. We discuss the results in comparison with analytic
% expressions for diffusion on general sparse graphs
% \cite{bray,samukhin:08} which independently derive stretched
% exponential relaxation expressions with the specific value
% $\beta=1/3$ for the exponent.
We argue that the KWW relaxation observed phenomenologically in
numerous complex systems just above their respective critical
temperatures is not an artifact, but is the signature of a universal
form of coarse grained configuration space morphology which precedes a
glass transition.

\section*{Laplace transforms and random networks}

Quite generally, any relaxation function $q(t)$ can equivalently be
characterized by its Laplacian, a relaxation mode density (or
eigenvalue density) function $\rho(s)$ defined by:
\begin{equation}
  \label{eq:relaxationfunction}
  q(t) \equiv \int_0^\infty \rho(s)e^{-s t}ds
\end{equation}
with the normalization condition
\begin{equation}
  \int_0^\infty \rho(s)ds =1
\end{equation}
In model systems it can be possible to establish analytically or
numerically the distribution $\rho(s)$ which can then be
inverted to obtain $q(t)$. The inverse Laplace transform of a
numerical or experimental $q_{K}(t)$ to obtain $\rho(s)$ is much more
difficult unless $q(t)$ is known to very high precision over a wide
range of $t$. This is an ill-conditioned problem as different
$\rho(s)$ distributions can lead to almost indistinguishable $q(t)$.

Pollard \cite{pollard} (see Berberan-Santos \cite{berberan2008})
provided an exact inversion of the pure stretched exponential
relaxation function $q_K(t)$ \ref{kohl} :
\begin{equation}
  \label{eq:laplace}
  \rho_{K,\beta}(s)=\frac{\tau}{\pi}\int_0^\infty \exp\left[ -u^\beta \cos\left(
      \frac{\beta \pi}{2}
    \right)
  \right] \cos\left[u^\beta\sin\left(
      \frac{\beta \pi}{2}
    \right)
  \right]
  \cos(s\tau u)
  \;du
\end{equation}
For $\beta < 1$, $\rho_{K,\beta}(s)$ can be expressed in terms
of elementary functions only for $\beta = 1/2$ \cite{pollard}; in that
case
\begin{equation}
  \label{eq:half}
  \rho_{K,1/2}(s)= [\tau/2\pi^{1/2}(s\tau)^{3/2}]\exp(-1/4s\tau)]
\end{equation}
To a good approximation, for general $\beta$ the large $s$ (short
time) limit takes the form $\rho_{K,\beta} \sim s^{-(1+\beta)}$
and the small $s$ (long time) limit the form $\rho_{K,\beta} \sim
(s)\exp[s^{-\beta/(1-\beta)}]$.

It should be kept in mind that at short times observed relaxation
functions usually deviate from the ``asymptotic'' form. Also at very
long times for finite sized systems the relaxation is controlled by
the smallest non-zero value of $s$, $s_{1}$. For time $t >
s_{1}^{-1}$ the relaxation will tend to a pure exponential, $q(t)
\sim \exp[-t s_{1}]$, but for large systems this condition corresponds
to extremely long times and we will not consider it.  What we are
interested in is to establish the form of the relaxation in the regime
where the mode distribution is no longer affected by short time
effects and where $\rho(s)$ can be considered continuous.

\section*{Random networks}

Random walks on the diluted $N$-simplex or hypertetrahedron which is
an Erd\"os-R\'enyi graph having dead ends and vertices with two
connections, was studied theoretically by Bray and Rodgers
\cite{bray1988} using Replica theory. They showed that
% at $p\rightarrow p_c$
in this model the return function $p_{ret}(t)$, the probability that
the walker will have returned to the origin after $t$ steps, behaves
like a stretched exponential with exponent $1/3$.

Samukhin {\it et al} \cite{samukhin2008} have made analytic studies of
random walks and relaxation processes on uncorrelated Random
Networks. They considered a stochastic process governed by the
Laplacian operator occurring on a random graph with $N^{*}$ nodes,
taking the limit as $N^{*} \to \infty$. They find that the determining
parameter in this problem is the minimum degree $q_{m}$ of vertices
(i.e. the minimum number of neighbors to any given vertex). For $q_m =
2$, meaning that the network is ``sparse'', the graph tends to a
random Bethe lattice in which almost all finite subgraphs are trees,
i.e., they contain almost no closed loops. In the present context the
essential statement of Samukhin {\it et al} \cite{samukhin2008} is
that when $q_m = 2$ the mode density function $\rho_{S}(s)$ for
this very general model can be approximated by
\begin{equation}
  \label{eq:laplacian}
  \rho_{S}(s) = s^{-4/3}\exp(-a/\sqrt{s})
\end{equation}
where
\begin{equation}
  \label{eq:defa}
  a=\sqrt{\frac{4\tau^-1}{3}}
\end{equation}
with a similar expression for $q_{m} = 1$ (graphs with dead
ends). Then for a graph with $N^{*}$ vertices the asymptotics at $t >
\ln N^{*}$ for the probability of return to the starting point at time
$t$ during a random walk on the network (the "autocorrelator"
\cite{samukhin2008}) will be
\begin{equation}
  \label{pretnet}
  p_{ret,S}(t) \sim t^{\eta}\exp[-3(a/2)^{2/3}t^{1/3}],
\end{equation}
a stretched exponential having exponent $1/3$, multiplied by a mildly
time dependent prefactor ($\eta$ is small). This limit should be
observable if the network size satisfies $(\ln N^{*})^{2/3} \gg 1$.

\section{Hypercube model}

We have already addressed the hypercube problem numerically through
Monte Carlo techniques \cite{lemke1996} and through the explicit
solution of Master equations \cite{lemke2000,almeida2000}.  In this
paper we extend these results by investigating the time evolution for
the autocorrelation function $q(N,t)$, the return probability
$p_{ret}(N,t)$, and the eigenvalue spectrum $\rho(N,s)$ for
diffusion on diluted hypercubes of dimension $N$ near the critical
occupation probability $p_c(N)$, for $N$ up to $28$.

% The estimation of the form of the eigenvalue spectrum is numerically
% the most demanding. We show that below a cut-off corresponding to
% the onset of the asymptotic the numerically evaluated relaxation
% mode spectra are consistent with the stretched exponential mode
% density form $\rho_{\beta}(\lambda)$ \ref{eq:laplace} with $\beta =
% 1/3$, or alternatively with the theoretical $q_{m}=2$ network mode
% density expression \ref{eq:laplacian} \cite{samukhin2008}. The
% long-time evolution of the return probability $p_{ret}(N,t)$ and of
% the autocorrelation function $q(N,t)$ (defined below) indeed both
% show stretched exponential behavior with exponent $\beta \sim 1/3$
% for all $N$.

% \section*{Definition of the model}

Consider a hypercube (or n-cube) in [high] dimension $N$,
$\mathbb{Q}_{N}$, with a fraction $p$ of its $2^N$ vertices occupied
at random. It is well established \cite{erdos1979,bollobas,borgs} that
there is a critical threshold at $p_{c}(N) \sim 1/N$. For $p > p_c(N)$
the occupied vertices having one or more occupied vertices as
neighbors make up a giant spanning cluster; for $p<p_c$ there exist
only small clusters (each with less than $N$ elements). By analogy
with the equivalent situation in randomly occupied Euclidean space we
will refer to $p_{c}$ as the ``percolation'' threshold.

Gaunt and Brak \cite{gaunt1984} predict that the dependency of the
critical site percolation concentration $p_c$ on a hypercubic lattice
of dimension $d$, $\mathbb{Z}^d$, or on a hypercube of dimension $N$,
$\mathbb{Q}_N$, is given to order $4$ by:
\begin{equation}
  p_c(\sigma) =\sigma+\frac{3}{2}\sigma^2+\frac{15}{14}\sigma^3+\frac{83}{4}\sigma^4\ldots
  \label{eq:pc}
\end{equation}
where $\sigma(d)=1/(2d -1)$ for the hypercubic lattice and
$\sigma(N)=1/(N-1)$ for the hypercube \cite{gaunt1976}. Although the
terms in this expression are expected to be exact, the demonstration
is not entirely rigorous \cite{gaunt1984}, and the series is obviously
truncated. Grassberger \cite{grassberger2003} tested the equation
(\ref{eq:pc}) through large scale Monte Carlo simulations on
$\mathbb{Z}^d$ and verified that for $d > 10$ it represents the
numerically determined $p_{c}(d)$ to within a small correction term.
We will work with samples having vertex concentrations $p(N)$ equal to
the values $p_c(N)$ given by the truncated series equation
(\ref{eq:pc}). For different samples $k$ the individual critical values
$p_c(k)$ will in fact be distributed about the average value
\cite{borgs}.

For $p > p_c(N)$ we can define a random walk along edges on the giant
cluster. Start at any vertex $i$ on the giant cluster.  Choose at
random a vertex $j$ on the hypercube, near neighbor to $i$.  If the
vertex $j$ is also on the giant cluster and so accessible, move to
$j$; otherwise the walker remains one time step longer at the vertex
$i$.  This evolution rule is chosen to mimic Monte Carlo simulations
using Metropolis dynamics.

We can compare the autocorrelation function $q(N,t)$ obtained from
this procedure, ($q(N,t)$ is defined in Eq. (\ref{eq:correlation})
below), to the time dependent autocorrelation $\langle
S_i(t).S_i(0)\rangle$ measured in thermodynamic models for systems of
Ising spins $S_{i}$ \cite{ogi1985} and even to experimental
magnetization decay results.  From a theoretical point of view it is
often more convenient to investigate the ``return probability''
$p_{ret}(t)$ that is basically the probability of finding the walker
at the origin of the system after $t$ steps ($p_{ret}(N,t)$ is defined
in Eq. (\ref{eq:pret}) below). For any network $p_{ret}(t)$ can be
defined, while $q(t)$ can be defined conveniently only on models such
as the hypercube which have a suitable metric.

The numerical data near criticality show that the long time
relaxations of the autocorrelation parameter $q(N,t)$ and of the
return probability $p_{ret}(N,t)$ are consistent with stretched
exponentials having an exponent $\beta = 1/3$ over many orders of
magnitude in time.

\section*{Algorithm}

The time evolution of the entire probability distribution for the
walker after $t$ steps, $\vec{\Pi}(t)$, can be described by a Master
Equation. At $t=0$ the walker is localized on a single vertex $i_o$ on
the hypercube; the probability distribution then diffuses over the
system at each time step following the equation:
\begin{equation}
  \label{eq:master}
  \Pi_i(t)   =\Pi_i(t-1) +
  \left[ \sum_{j}
    \Pi_i(t-1)W(j\to i)-
    \Pi_j(t-1)W(i\to j) \right]
\end{equation}
where $W(i\to j)$ represents the transition probability that is given
by:
\begin{equation}
  \label{eq:transition}
  W(i\to j)=\left\{
    \begin{array}{l l}
      \frac{1}{N} & \mbox{if $i$ vertex  and $j$ vertex  are allowed} \\
      0 & \mbox{otherwise}
    \end{array}
  \right.
\end{equation}
The equation (\ref{eq:master}) can be rephrased as:
\begin{equation}
  \label{eq:operator}
  \vec{\Pi}(t)=F\vec{\Pi} (t-1)
\end{equation}
where $F$ is the linear evolution operator.

Since this process is Markovian we can diagonalize $F$; the smallest
eigenvalue corresponding to the infinite time equilibrium limit (where
all sites become equally populated) is 1. We can determine $U$ and $D$
satisfying:
\begin{equation}
  \label{eq:diag}
  F=U^TDU
\end{equation}
where $D$ is a diagonal matrix. For practical reasons it is convenient
to diagonalize $F$ so as to investigate the temporal evolution of the
relevant quantities. We use:
\begin{equation}
  \label{eq:time-evol}
  \Pi(t)=F^t\Pi(0)=U^TD^tU\Pi(0)
\end{equation}
We choose the initial condition as:
\begin{equation}
  \label{eq:initial}
  \Pi_i(0)=\delta_{ii_o}
\end{equation}
where $i_o$ is a vertex on the giant cluster.

The value of the normalized autocorrelation function $q(t)$ after time
$t$ for a given walk starting from $i_o$ and arriving at $i$ after
time $t$ can be defined by:
\begin{equation}
  \label{eq:correlation}
  q(t)  =\left\langle \frac{1}{N_G}\sum_{i_o} \sum_{i} \Pi_i(t)
    \frac{N-2d_H(i,i_o)}{N}- q_\infty \right\rangle
\end{equation}
where $d_H(i,i_o)$ is the Hamming distance between vertex $i$ and the
initial state, $N_G$ is the number of vertices on the giant cluster,
$q_{t=\infty}$ for a given realization is given by:
\begin{equation}
  \label{eq:correlation2}
  q_{t=\infty}  =\frac{1}{N_G^2}\sum_{ii_o}
  \frac{N-2d_H(i,i_o)}{N}
\end{equation}
and the averages are over different realizations of the diluted
hypercube.

We also calculated $p_{ret}$ defined by:
\begin{equation}
  \label{eq:pret}
  p_{ret}(t)=\left\langle \frac{1}{N_G}\sum_{i_o} \Pi_{i_o}(t)-\frac{1}{N_G}\right\rangle
\end{equation}
We can show that:
\begin{equation}
  \label{eq:pret2}
  p_{ret}(t)=\frac{1}{N_G}\left\langle \sum_{j}\lambda_j^t
    -\frac{1}{N_G} \right\rangle
\end{equation}
% {\bf I suppose \lambda_j^t means (\lambda_j)^t ?}
This quantity is easier to calculate theoretically than $q(t)$, but it
is not useful to compare with results on model spin systems or
experiments. We can write this equation in a more convenient form:
\begin{equation}
  \label{eq:pret3}
  p_{ret}(t)=\frac{1}{N_G}\left\langle \sum^\prime_{s\neq 0}e^{-s_it}\right\rangle
\end{equation}
where $s=-\ln \lambda$ and we excluded $\lambda =1$ eigenvalue. Another
convenient form for investigating $p_{ret}$ is:
\begin{equation}
  \label{eq:dens-exp}
  p_{ret}(t)=\int_0^\infty ds\rho(s) e^{-ts} 
\end{equation}
where the density $\rho$ is defined by:
\begin{equation}
  \label{eq:density}
  \rho(\lambda)=\left\langle \frac{1}{N_G-1} \sum_i \delta(s-s_i )
  \right\rangle
\end{equation}

Our numerical workflow can be summarized as follows:
\begin{enumerate}
\item generation of a diluted hypercube
\item determination of the giant cluster
\item determination of the eigenvalues and eigenvectors of $F$
\item calculation of $\rho(s)$, $q(t)$ and $p_{ret}(t)$
\end{enumerate}

The algorithm was implemented on Mathematica 8.0 and the simulations
were performed on a Intel Xeon 2.27 Ghz with 24 Gbytes of Ram Memory.
A single simulation for the $N=28$ cost 12 hours.  The algorithm
demands 24 Gbytes of memory for this case.

Calculations were made with hypercubes of dimension $N=10, 12, 14, 16,
18, 20, 22, 24, 26$ and $28$.  All the calculations were performed at
$p_c(N)$ values given by equation (\ref{eq:pc}); this condition is
important since it allows us to scale conveniently data for systems
having different dimensions $N$. It is useful to be able to include
data for smaller $N$ in the global analysis as in these samples we
deal with much smaller matrices which is simpler computationally.

All vertices on the giant cluster were used as starting points, except
for the largest systems $N=26$ and $28$ where we have not used all
possible initial states $i_o$. For these sizes we approximated $q(t)$
and $p_{ret}(t)$ by using only $1000$ randomly chosen initial states
for each realization. We have tested the accuracy of this
approximation and we concluded that the error was very small (even for
the smaller sizes).  We studied $1000$ different realizations of the
hypercube for all sizes $N$ except for $N=28$ when we have studied
$100$.

\section{Numerical data}

\begin{figure}
  \begin{center}
    \includegraphics[scale=0.6]{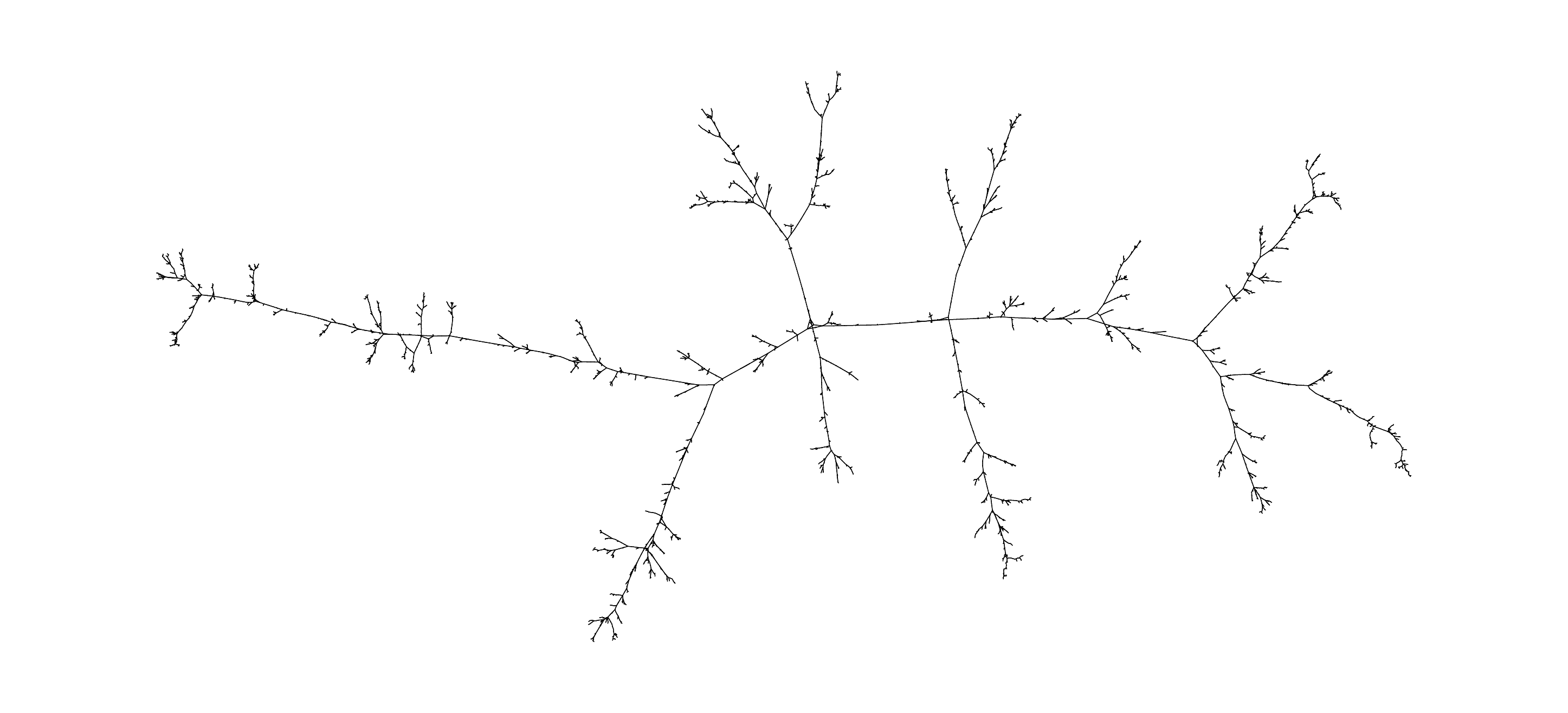}
  \end{center}
  \caption{ A graphical representation of a diluted $\mathbb{Q}_{24}$
    exactly at $p_c$. The picture shows that the network presents no
    loops.}
    \label{fig:net}
  \end{figure}

On Figure (\ref{fig:net}) we represent a graphical representation of a
diluted $\mathbb{Q}_{24}$ for this particular sample the graph is a tree
showing the validity of the approximation proposed by \cite{samukhin2008}. 

\begin{figure}
  \begin{center}
    \includegraphics[scale=0.8]{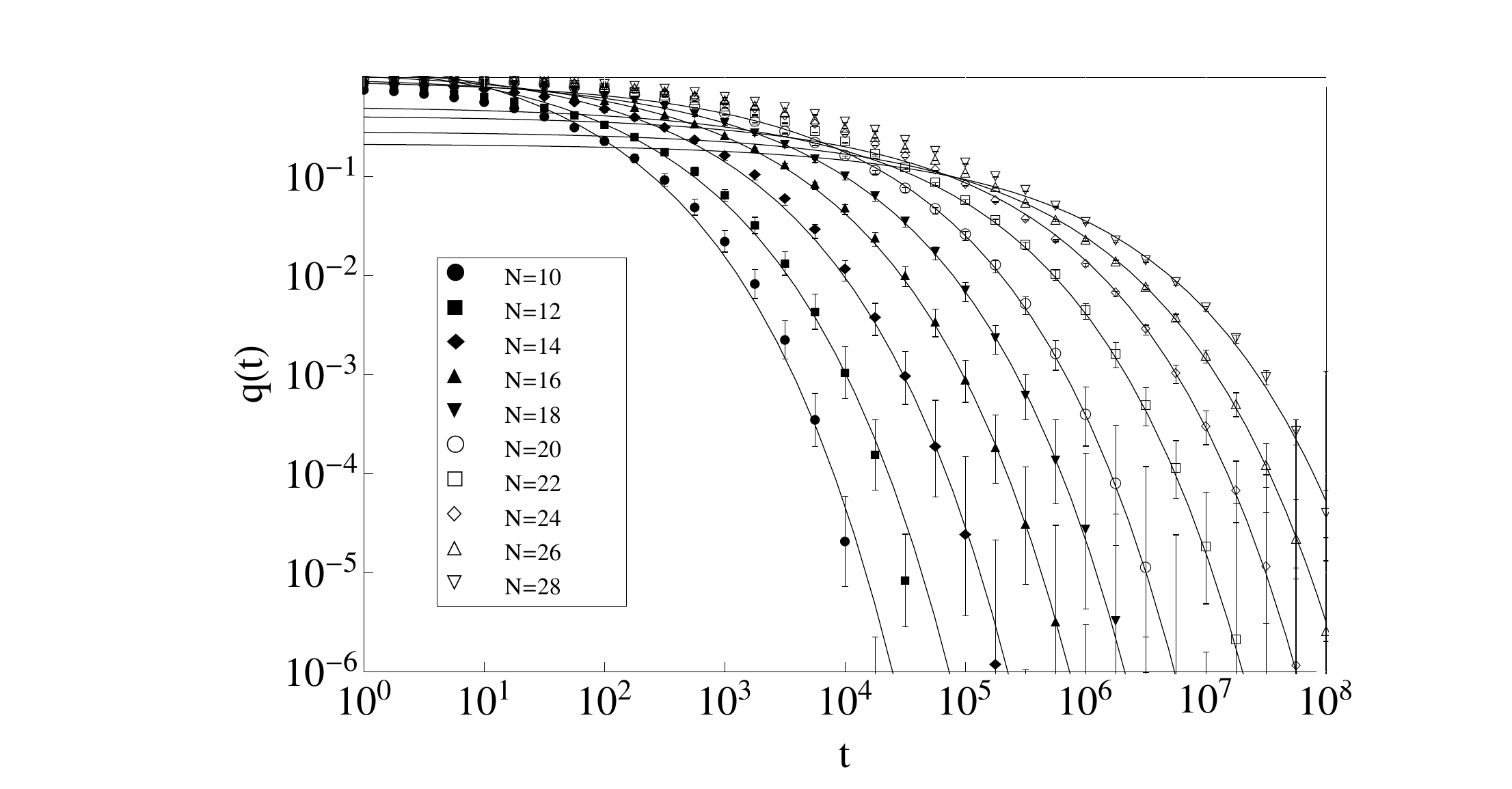}
  \end{center}
  \caption{ The relaxation of the autocorrelation function $\log
    q(N,t)$. Eqn.(\ref{eq:correlation}), against $\log(t)$ for 
$N$ from $10$ to $28$. .
% The color    code indicates the hypercube dimension $N$ from $10$ to $28$. 
}
  \label{fig:corrlog}
\end{figure}

\begin{figure}
  \begin{center}
    \includegraphics[scale=0.6]{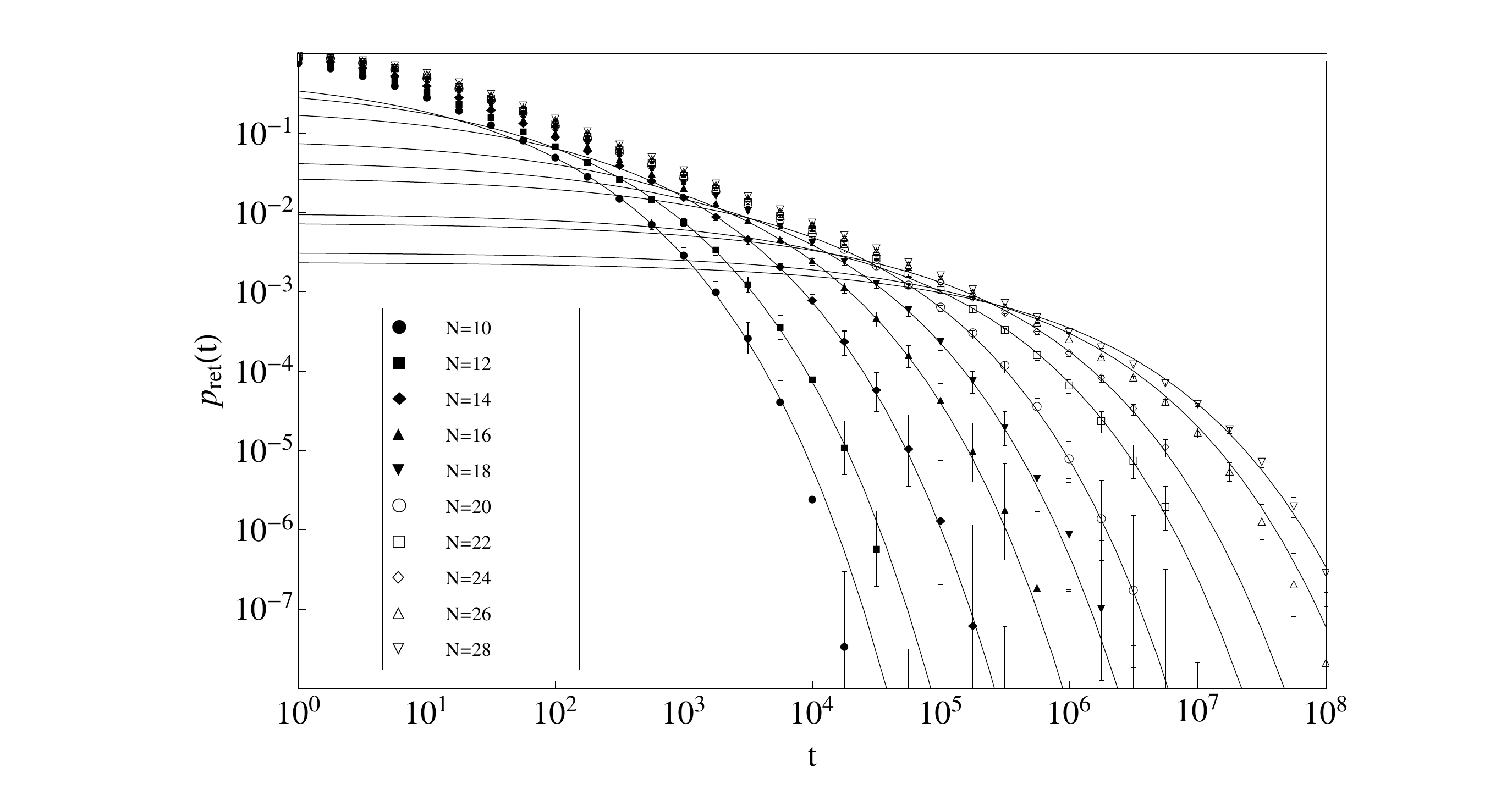}
  \end{center}
  \caption{ The decay of the return probability $\log p_{ret}(N,t)$,
    Eqn.(\ref{eq:pret}), against $\log t$ for $N$ from $10$ to $28$. 
%The color code indicates the   hypercube dimension $N$ from $10$ to $28$.
  }
  \label{fig:pretlog}
\end{figure}

  The time evolution for the autocorrelation functions $q(N,t)$
  (\ref{eq:correlation}) is depicted in Figure \ref{fig:corrlog} against
  $\log(t)$.  On Figure \ref{fig:pretlog} we show the equivalent
  results for the return probability $p_{ret}(N,t)$.

  In all cases we have fitted the long time part of the curves using
  the expression:

\begin{equation}
  f(t)=A\exp\left[- \left( \frac{t}{\tau}\right)^{1/3} \right]
\end{equation}

\begin{figure}
  \begin{center}
    \includegraphics[scale=0.8]{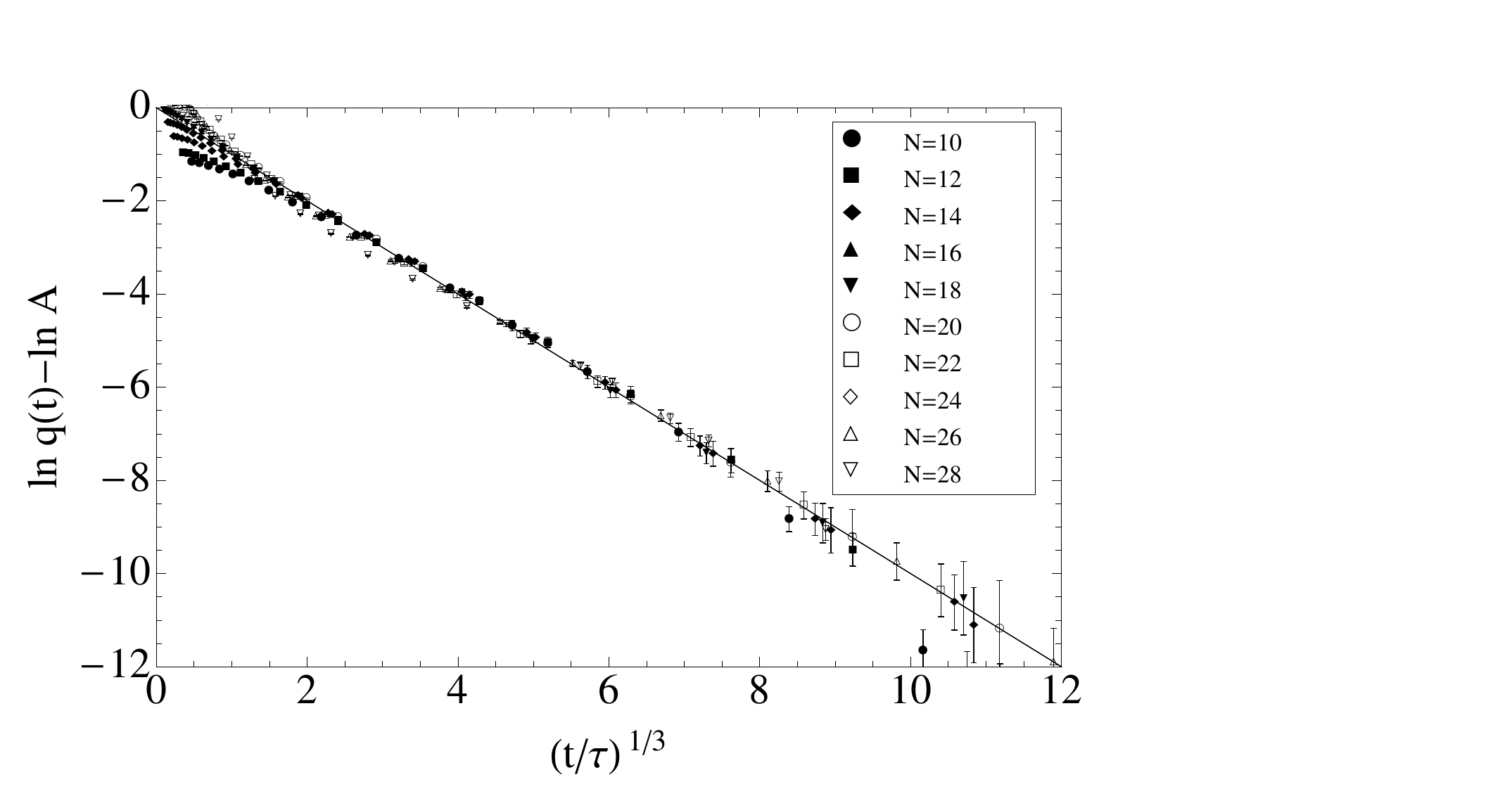}
  \end{center}
  \caption{ The decay of the normalized autocorrelation function $\ln
    (q(N,t)/A_{q}(N))$ against $(t/\tau)^{1/3}$. For stretched
    exponentials with exponent $\beta=1/3$ in the long time regime the
    data should lie on a straight line in this form of plot, as
    observed. }
  \label{fig:corrt}
\end{figure}

\begin{figure}
  \begin{center}
    \includegraphics[scale=1.0]{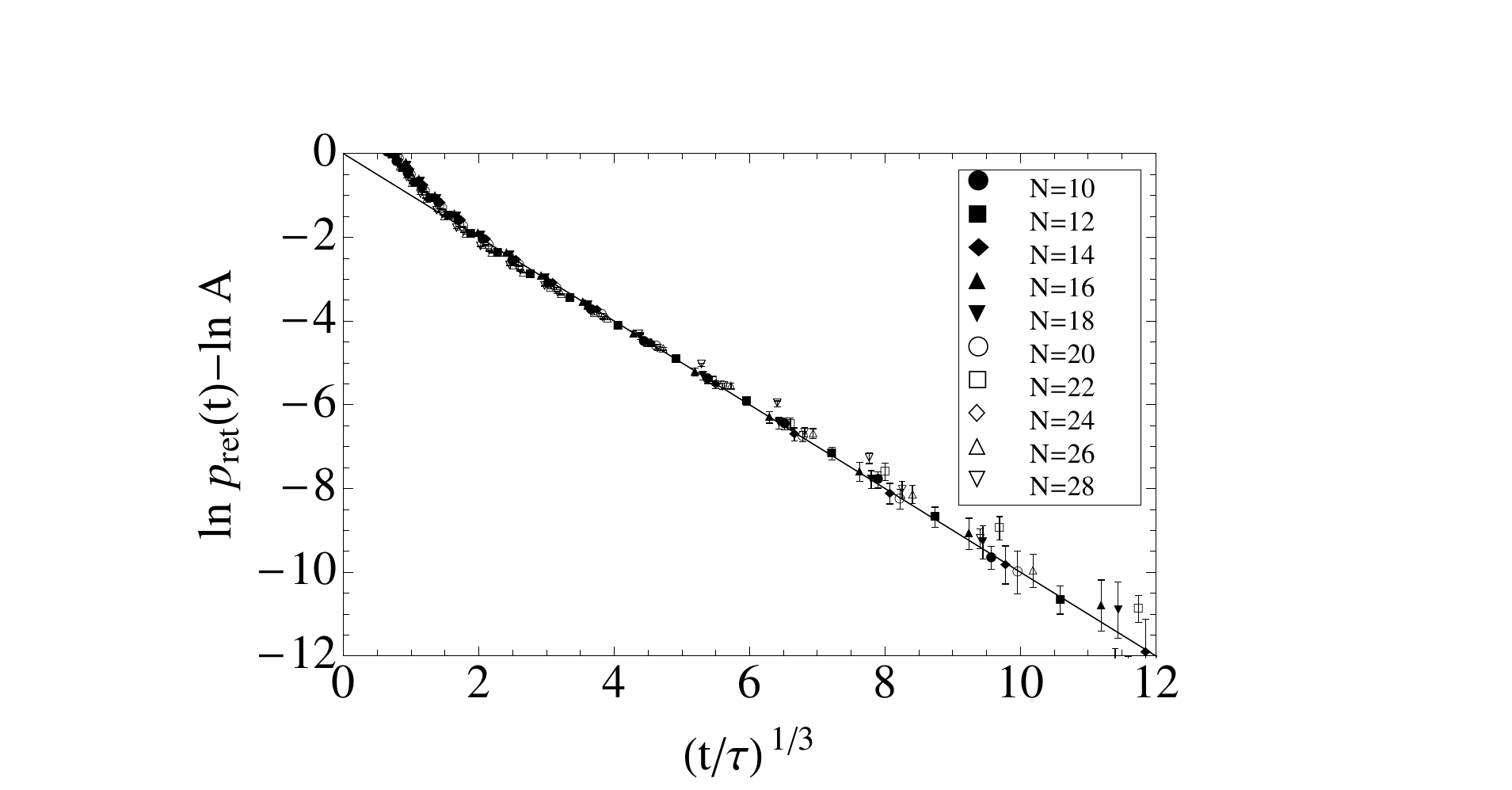}
  \end{center}
  \caption{ The decay of the normalized return probability $\log
    (p_{ret}(N,t)/A_{ret}(N))$ against $(t/\tau(N))^{1/3}$. For
    stretched exponentials with exponent $\beta=1/3$ in the long time
    regime the data should lie on a straight line on this form of
    plot, as observed. }
  \label{fig:prett}
\end{figure}

In Figures \ref{fig:corrt} and \ref{fig:prett} we present the same
results in a different manner so as to demonstrate the stretched
exponential long time behavior. On the $x$ axis the time scale is
normalized with $x(t) = (t/\tau(N))^{1/3}$ and on the $y$ axis the
measured $q(N,t)$ or $p_{ret}(N,t)$ are normalized so $y(N,t) = \ln
(q(N,t)/A_{q}(N))$ and $y(N,t) = \ln (p_{ret}(N,t)/A_{ret}(N))$
respectively.  In these plots a stretched exponential with exponent
$1/3$ is a straight line as observed; we have chosen the normalization
factors $\tau(N)$ and $A_{q}(N), A_{ret}(N)$ so that data for
different hypercube dimensions $N$ collapse. This form of plot allows
one to distinguish clearly between the short time regime and the
stretched exponential regime; the latter can be seen to extend over a
wide time range until measurements are limited by the statistical
noise. The effective exponent $\beta = 1/3$ is independent of $N$ to
within the statistics.

\begin{figure}
  \begin{center}
   \includegraphics[scale=0.25]{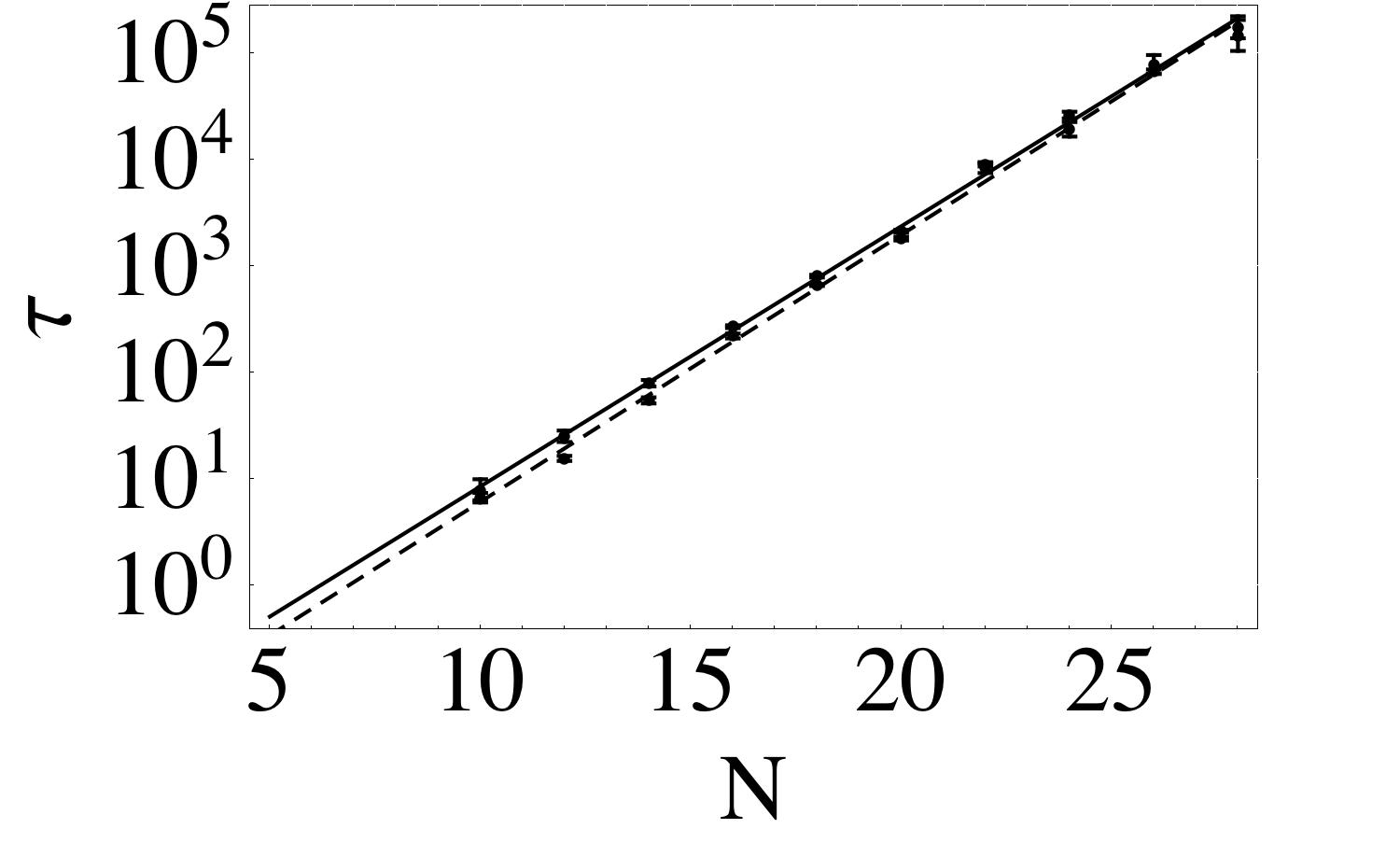}
  \end{center}
  \caption{ The dependence of the time scale $\tau(N)$ with dimension
    for the return probability $p_{ret}(N,t)$ (in red) and
    autocorrelation $q(N,t)$ (in blue). }
  \label{fig:tau}
\end{figure}

On Figure \ref{fig:tau} we show the size dependence on the $\tau(N)$
time scale parameter from the fits of the autocorrelation $q(t,N)$ and
the return probability $p_{ret}(t,N)$ data. The data can be fitted by
fitted by
\begin{equation}
  \tau(N)=B 10^{\gamma N}
  \label{eq:tau}
\end{equation}
with the fit parameters $\gamma = 0.24\pm 0.1$ and $B= 1.5\pm 0.1$ for
autocorrelation function, $\gamma = 0.24\pm 0.05$ and $B= 1.7\pm 0.2$
for the return probability. The values of the time scaling parameters 
$\tau(N)$ for the two
different observables are identical within the precision of the
measurements.

The most fundamental way to understand the system dynamics is through
investigating the eigenvalue spectra; the stretched exponential long
time behavior depends exclusively on the density of the eigenvalues
above the smallest eigenvalue, in the region where the distribution
for a finite size sample can still be considered to be continuous. A
given spectrum leads unambiguously to a unique relaxation function,
while it is much more difficult to determine the precise form of a
mode spectrum from a relaxation function.

\begin{figure}
  \begin{center}
    \includegraphics[scale=0.4]{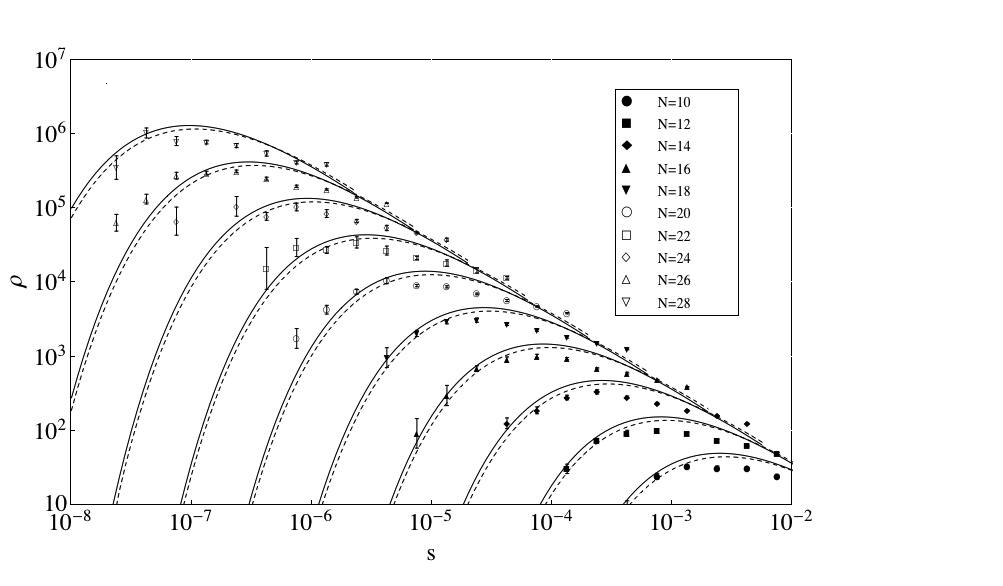}
  \end{center}
  \caption{Spectral density data $\rho(N,s)$ from the hypercube
    evaluations together with the exact Laplace transforms
    $\rho_{1/3}(s)$ (\ref{eq:laplace}) for stretched exponentials with
    $\beta=1/3$ and $\tau(N)$ values equal to the numerical estimates
    \ref{eq:tau} (dashed lines), and the analytic sparse network
    expression (\ref{eq:laplacian}),\cite{samukhin2008} (full
    lines). The values are normalized (see text). }
  \label{fig:rho}
\end{figure}

On Figure \ref{fig:rho} we compare the mode density $\rho(s)$ obtained
through the present simulations with the theoretical expressions. All
the numerical results were obtained using $1000$ different
realizations of the diluted hypercube at each dimension
$N$. Unfortunately in practice the calculations of $\rho(s)$ are
numerically demanding because of strong sample to sample
fluctuations. The spectra were first binned in the form of histograms.
% The numerical distributions were normalized.
We defined a cut-off $\lambda_{min}(N)$ or equivalently
$s_{max}(N)=-\ln \lambda_{min}(N)$ to eliminate the short time effects
and selected the eigenvalues on the interval $s\in (0,
s_{max}(N))$.  We choose  $s_{max}=2/\tau(N)$ for all dimensions. 
We divided this interval in bins equally spaced on a
logarithmic scale and then calculated the densities for each interval,
normalizing the frequencies by the length of the intervals.

The continuous curves were calculated from the expression (\ref{eq:laplace})
for $\rho_{K,1/3}(\lambda)$  and from the approximate
analytic expression (\ref{eq:laplacian}) for $\rho_{S}(\lambda)$  using
$\tau(N)$ estimated from equation (\ref{eq:tau}). To compare with
simulation results we normalized $\rho(\lambda)$ functions using:

\begin{equation}
  C^{-1}=\int_{0}^{s_{max}} \rho(s)ds
\end{equation}
and
\begin{equation}
  \rho^\prime(s)=C\rho(s)
\end{equation}

Over the ranges for which reliable data points have been obtained the
measured mode spectrum densities $\rho(N,s)$ closely resemble the
corresponding parts of the calculated spectra from the Laplace
transform $\rho_{K,1/3}(s)$, (\ref{eq:laplace}) or the analytic
$\rho_{S}(s)$ spectrum (\ref{eq:laplacian}) \cite{samukhin2008} (which
are in fact very similar to each other). The numerical spectra for the
hypercube model are indeed consistent with the mode density spectral
form derived analytically for the more general random network model
\cite{samukhin2008}.

\section*{Discussion and conclusions}

We have studied numerically relaxation through random walks along near
neighbor edges on the giant cluster of vertices in randomly diluted
hypercubes of dimensions up to $N=28$ near the percolation threshold
for the cluster. The data show clearly that at the percolation
threshold concentration $p_c(N)$, the relaxation mode spectrum, the
time dependence of the autocorrelation $q(N,t)$, and the return
probability $p_{ret}(N,t)$, are all consistent with asymptotic
stretched exponential relaxation $\exp[-(t/\tau(N))^\beta]$ having
exponent $\beta = 1/3$. The time scale $\tau(N)$ increases
exponentially with dimension $N$, Eqn. (\ref{eq:tau}). The observed
eigenvalue spectra demonstrate that the dynamical $q(N,t)$ behavior
previously obtained from Monte Carlo simulations and from numerical
solutions of the master equation
\cite{lemke1996,lemke2000,almeida2000} does not represent a crossover
between different exponential regimes, but that it is the consequence
of a specific wide eigenvalue spectrum.

A final long time crossover to a pure exponential (which would
correspond to a regime where the effective relaxation mode spectrum is
reduced to a gap between the ground state and the lowest mode) is not
visible in the data.

This diluted hypercube model at threshold can be considered as the
limiting high dimensional case of percolation on sphere-like
spaces. Alternatively it can be considered as a specific explicit
example of a generic sparse random network.  The observed stretched
exponential behavior with exponent $\beta =1/3$ on the dilute
hypercube at the percolation threshold is consistent with the
predictions of the sphere-like percolation approach \cite{ian1} and
with studies of random walks on sparse random networks
\cite{bray1988,samukhin2008}, where the same stretched exponential
relaxation with the same exponent $\beta = 1/3$ has been derived
analytically.

For a physical system, configuration space can be imagined as a very
high dimensional graph. The system's dynamics is equivalent to a
random walk of the point representing the instantaneous state of the
system among those vertices of the graph which are thermodynamically
accessible.  We suggest that when the stretched exponential
$\exp[-(t/\tau)^{1/3}]$ form of limiting relaxation with diverging
$\tau$ is observed numerically or experimentally for the
autocorrelation function relaxation $q(t)$ in complex physical systems
(which is often the case, see for instance
\cite{ogi1985,angelani,billoire})it is the signature of a
configuration space tending to a percolation threshold and having a
sparse random network topology.

\section*{Acknowledgements}

This work was supported by FAPESP grant no. 09/10382-2. This research
was supported by resources supplied by the Center for Scientific
Computing (NCC/GridUNESP) of the S\~ao Paulo State University (UNESP).

% \section*{References}

\end{document}